\documentclass[prl,twocolumn,showpacs,preprintnumbers,]{revtex4}
\usepackage{graphicx}
\usepackage{natbib}
\usepackage{latexsym}

\begin{document}

\title{Anisotropic solitons in dipolar Bose-Einstein Condensates}
\author{I. Tikhonenkov$^a$, B. A. Malomed$^b$, and A. Vardi$^a$}
\affiliation{$^a$Department of Chemistry, Ben-Gurion University of the Negev, P.O.B. 653,
Beer-Sheva 84105, Israel}
\affiliation{$^b$Department of Physical Electronics, School of Electrical Engineering,
Faculty of Engineering, Tel Aviv University, Tel Aviv 69978, Israel}

\begin{abstract}
Starting with a Gaussian variational ansatz, we predict anisotropic bright
solitons in quasi-2D Bose-Einstein condensates consisting of atoms with
dipole moments polarized \emph{perpendicular} to the confinement direction.
Unlike isotropic solitons predicted for the moments aligned with the
confinement axis [Phys. Rev. Lett. \textbf{95}, 200404 (2005)], no sign
reversal of the dipole-dipole interaction is necessary to support the
solitons. Direct 3D simulations confirm their stability.
\end{abstract}

\maketitle

The creation of two- and three-dimensional (2D and 3D) solitons is a challenge
to experiment in optical and atomic physics \cite{review}. Both
\textquotedblright light bullets" in nonlinear optics \cite{Silberberg} and
2D/3D matter-wave solitons are objects of profound significance. In optics,
the best result was the creation of partially localized (quasi-2D)
spatiotemporal solitons in crystals with quadratic nonlinearity \cite%
{Frank}. As concerns matter waves, quasi-1D bright solitons were
demonstrated in self-attractive Bose-Einstein condensates (BECs) of $^{7}$Li
and $^{85}$Rb atoms \cite{Li-Rb}, and gap solitons were created in a
self-repulsive $^{87}$Rb condensate trapped in an optical lattice (OL) \cite%
{Markus}. A fundamental impediment to the creation of 2D and 3D bright
solitons with attractive cubic nonlinearity is their instability to
collapse. Theoretically considered stabilization schemes include the use of
OLs \cite{BBB} and periodic time modulation of the nonlinearity by means of
a Feshbach resonance (FR) in ac magnetic field \cite{Feshbach}. Another
approach assumes using 2D or 3D OLs to create gap solitons in respective
settings with repulsive nonlinearity \cite{GS}. So far, however, no
experimental realization of matter-wave solitons in a 2D or 3D setting has
been reported, and other stabilization mechanisms are highly desirable.

One such mechanism may utilize the nonlocal interaction between dipolar
atoms in the quantum Bose gas \cite{dipBEC}. Potential implementations
include BECs of magnetically polarized Cr atoms \cite{Cr}, permanent
electric dipoles in quantum gases of heteronuclear molecules\cite{hetmol},
and electric dipoles induced by laser illumination \cite{lightDD} or by
strong dc electric field. In this context, it is natural to consider a
nearly 2D \textquotedblright pancake" geometry, the simplest (isotropic)
configuration having dipole moments aligned along the tight-confinement
direction, denoted here as $z$. This approach to the creation of quasi-2D
solitons in dipolar BECs was elaborated in Ref. \cite{Pedri05}, using a
variational approximation based on the Gaussian ansatz 
$\psi _{\mathrm{iso}}=\pi
^{-3/4}\alpha ^{1/2}\gamma ^{1/4}\exp \left[ -\left( \alpha \left(
x^{2}+y^{2}\right) +\gamma z^{2}\right) /2\right] $, where $\alpha $ and $%
\gamma $ are the inverse squares of the radial and axial widths,
respectively. This trial-function generates the following Gross-Pitaevskii (GP)
energy functional,
\begin{equation}
E\left\{ \psi _{\mathrm{iso}}\right\} =\frac{1}{2}\left( \alpha +\frac{%
\gamma }{2}+\frac{1}{2\gamma }\right) +\alpha \sqrt{\frac{\gamma }{2\pi }}%
\left( \frac{g}{4\pi }+\frac{g_{d}}{3}f(\kappa )\right) ,  \label{enrgperp}
\end{equation}%
where $\kappa \equiv \sqrt{\gamma /\alpha }$ is the aspect ratio, and
\begin{equation}
f(\kappa )\equiv \frac{2\kappa ^{2}+1}{\kappa ^{2}-1}-\frac{3\kappa ^{2}}{%
\left( \kappa ^{2}-1\right) \sqrt{\left\vert \kappa ^{2}-1\right\vert }}%
\left\{
\begin{array}{l}
\arctan \sqrt{\kappa ^{2}-1},~\kappa >1 \\
\frac{1}{2}\ln \left( \frac{1+\sqrt{1-\kappa ^{2}}}{1-\sqrt{1-\kappa ^{2}}}\right)
,~\kappa <1%
\end{array}%
\right. ~.  \label{f}
\end{equation}%
Here and below, length, time, and energy are scaled as $\mathbf{r}%
\rightarrow \mathbf{r}/l_{\perp }$, $t\rightarrow \omega _{\perp }t$, and $%
E\rightarrow E/\left( \hbar \omega _{\perp }\right) $, where $\omega _{\perp
}$ is the trap frequency in the $z$ direction, and $l_{\perp }\equiv \sqrt{%
\hbar /m\omega _{\perp }}$. The interaction strengths are $g\equiv 4\pi
Na_{s}/l_{\perp }$ and $g_{d}\equiv Nd^{2}m\left( \hbar ^{2}l_{\perp
}\right) ^{-1}$, where $a_{s}>0$ is the $s$-wave scattering length, $d$ and $%
m$ the atomic dipole moment and mass, $N$ the number of atoms, and the wave
function is subject to normalization $\int \left\vert \psi \left( \mathbf{r}%
\right) \right\vert ^{2}d\mathbf{r}=1$.

\begin{figure}[tbp]
\centering
\includegraphics[width=0.5\textwidth]{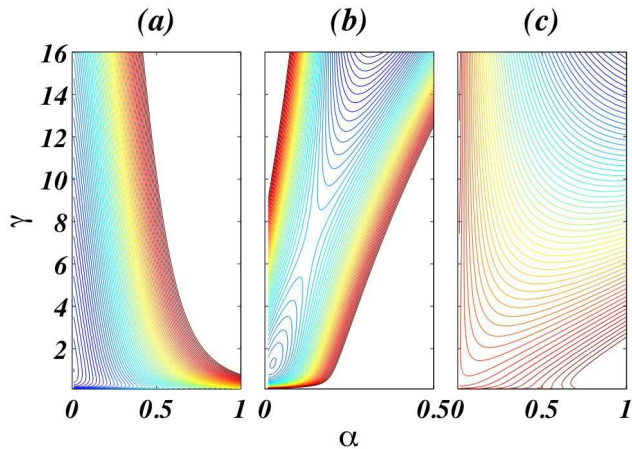}
\caption{(color online) Gross-Pitaevskii energy functional (\protect\ref%
{enrgperp}), as a function of variational parameters $\protect\alpha $ and $%
\protect\gamma $, with $g=500$ and (a) $g_{d}/g=-0.10$, (b) $g_{d}/g=-0.20$,
(c) $g_{d}/g=-0.40$. Weak dipolar interaction in (a) leads to expansion of
the condensate, whereas strong attraction in (c) results in collapse. For
intermediate strengths of the dipole interaction in (b), stable quasi-2D
solitons are possible.}
\label{fig1}
\end{figure}

For fixed $\gamma $ in the absence of dipole-dipole (DD) interactions ($%
g_{d}=0$), both kinetic and contact-interaction energies (the first and
fourth terms in Eq. (\ref{enrgperp})) scale as the inverse square of the
radial size in the 2D plane, resulting in the well-known instability of the
\textit{Townes solitons} \cite{Berge}. Stabilization is shown by fixing $%
\gamma =1$ in Eq. (\ref{enrgperp}) and searching for a minimum of the energy
as a function of $\alpha $, which results in the necessary condition \cite%
{Pedri05},
\begin{equation}
\left( 2/3\sqrt{2\pi }\right) g_{d}<1+g/(2\pi )^{3/2}<-\left( 4/3\sqrt{2\pi }%
\right) g_{d},  \label{condperp}
\end{equation}%
that may only hold for $g_{d}<0$. The interpretation of this condition
reveals the origin of self-trapping in the proposed configuration. Since the
DD interaction is repulsive and attractive, respectively, for
\textquotedblleft side-by-side" and \textquotedblleft head-to-tail" dipole
pairs, the overall dipolar energy varies from negative for a prolate
structure, with $L_{\rho }\sim \alpha ^{-1/2}\ll L_{z}\sim \gamma ^{-1/2}$,
through zero for the isotropic case, $L_{\rho }=L_{z}$, to positive for the
oblate condensate with $L_{\rho }\gg L_{z}$ (this variation is evident in
the sign change of function (\ref{f}) at $\kappa =1$). Fixing $L_{z}$, a
\emph{maximum} may be attained for the total energy as a function of $%
L_{\rho }$, provided that the DD term is sufficiently large to reverse the
sign of the sum of the kinetic and contact-interaction energies, all scaling
as $1/L_{\rho }^{2}$. This unstable maximum may be turned into a stable
minimum if the sign of the dipolar interaction strength is reversed to $%
g_{d}<0$. This goal may be attained by using rotating external fields \cite%
{Giovanazzi02}, but in combination with the necessity to reduce contact 
interactions in order to make the DD interaction dominant, it introduces 
an essential complication to the experiment \cite{Pedri05}, and it is not 
possible for electric moments induced by the polarizing dc field.%
%In the limit when the 2D kinetic energy is negligible in comparison to the contact-interaction strength, g/(2(2p)^{3/2})»1, condition (<ref>condperp</ref>) simplifies to |g_{d}/g|>3/8p.

Condition (\ref{condperp}) does not guarantee a true minimum of $E(\alpha
,\gamma )$ in the $\left( \alpha ,\gamma \right) $ plane. For weak DD
interactions, Eq. (\ref{condperp}) does not hold, and all fixed-$\gamma $
energy curves are monotonic in $\alpha $, as shown in Fig. \ref{fig1}(a).
For larger values of $|g_{d}|$, a shallow local minimum (elliptic point)
does appear, as in Fig. \ref{fig1}(b). It gives rise to a quasi-2D soliton,
which is isolated from collapsing in the direction of $\alpha ,\gamma
\rightarrow \infty $ by a saddle point. However, with further increase of
the strength of the DD interactions, the elliptic and saddle points merge
and disappear, resulting in an essentially 3D collapse, as seen in Fig. \ref%
{fig1}(c). This is the source of the \textquotedblleft breathing
instability" in Ref. \cite{Pedri05}.

Related works considered the stability of vortex lattices \cite{Pu} and
Thomas-Fermi-like trapped states \cite{Shai} in dipolar BECs in the same
oblate geometry. Also relevant are theoretically elaborated examples of
isotropic optical solitons supported by thermal and other nonlocal
nonlinearities in the 2D geometry \cite{Optics}.

In this work, we aim to explore a new setting with dipole moments polarized
in the 2D plane, i.e., \emph{perpendicular} to the tight-confinement axis,
which we now designate $y$, reserving label $z$ for the dipolar axis. With
this notation, the GP energy functional is
\begin{eqnarray}
E\left\{ \psi \right\}  &=&\frac{1}{2}\int \left( |\nabla \psi (\mathbf{r}%
)|^{2}+y^{2}|\psi (\mathbf{r})|^{2}+g|\psi (\mathbf{r})|^{4}\right) d\mathbf{%
r}  \label{Epar} \\
&&+\frac{g_{d}}{2}\int \int \left[ 1-\frac{3\left( z-z^{\prime }\right) ^{2}%
}{\left\vert \mathbf{r}-\mathbf{r^{\prime }}\right\vert ^{2}}\right] |\psi (%
\mathbf{r^{\prime }})|^{2}|\psi (\mathbf{r})|^{2}\frac{d\mathbf{r}d\mathbf{%
r^{\prime }}}{\left\vert \mathbf{r}-\mathbf{r^{\prime }}\right\vert ^{3}},
\nonumber
\end{eqnarray}%
Substituting the general anisotropic ansatz, $\psi _{\mathrm{aniso}}=\pi
^{-3/4}\left( \alpha \beta \gamma \right) ^{1/4}\exp \left[ -(1/2)\left(
\alpha x^{2}+\beta y^{2}+\gamma z^{2}\right) \right] $, into (\ref{Epar}),
we find (with $\kappa _{x}\equiv \sqrt{\gamma /\alpha }$, $\kappa _{y}\equiv
\sqrt{\gamma /\beta }$)
\begin{equation}
E(\psi _{\mathrm{aniso}})=\frac{1}{4}(\alpha +\beta +\gamma )+\frac{1}{%
4\beta }+\sqrt{\frac{\alpha \beta \gamma }{2\pi }}\left[ \frac{g}{4\pi }+%
\frac{g_{d}}{3}h(\kappa _{x},\kappa _{y})\right] ,  \label{enrgpar}
\end{equation}%
\begin{equation}
h(\kappa _{x},\kappa _{y})\equiv \int_{0}^{1}\frac{3\kappa _{x}\kappa
_{y}x^{2}dx}{\sqrt{1+(\kappa _{x}^{2}-1)x^{2}}\sqrt{1+(\kappa
_{y}^{2}-1)x^{2}}}-1.  \label{h}
\end{equation}

While $h(\kappa _{x},\kappa _{y})$ may be expressed in terms of elliptic
integrals, its behavior is more readily seen directly from Eq. (\ref{h}). We
first consider the variation of the characteristic width, $L_{z}=1/\sqrt{%
\gamma }$, along polarization axis $z$. At small $L_{z}$ ($\gamma \gg \alpha
,\beta $), we have $h(\kappa _{x},\kappa _{y})\rightarrow h(\infty ,\infty
)=2$. On the other hand, $h(\kappa _{x},\kappa _{y})\rightarrow h(0,0)=-1$
for $\gamma \ll \alpha ,\beta $. Thus for $g_{d}>0$, i.e., for the \emph{%
natural} sign of the DD interaction, Eq. (\ref{enrgpar}) demonstrates a
switch from repulsion at small $L_{z}$ to attraction at large $L_{z}$, thus
predicting a stable bound state as concerns the variation of $L_{z}$,
provided that $g_{d}$ is large enough.

While the above consideration makes stable self-trapping along the
polarization axis evident, behavior along the other unconfined direction, $x$%
, is more subtle. Indeed, one might expect expansion of the condensate in
this direction, as the side-by-side DD interaction is repulsive,  However,
examination of expression (\ref{h}) yields the asymptotic limits, $%
h(0,\kappa _{y})=-1$ and $h(\infty ,\kappa _{y})=\left( 2\kappa
_{y}-1\right) /\left( \kappa _{y}+1\right) $, so that for $g_{d}>0$ and $%
\kappa _{y}<1/2$, the DD interaction remains effectively \emph{attractive}
even for arbitrarily large values of $L_{x}$. For $\kappa _{y}>1/2$ (i.e.
for $L_{z}>2L_{y}$), the DD interaction changes from the attraction for $%
\kappa _{x}\rightarrow 0$ to repulsion at $\kappa _{x}\rightarrow \infty $.
Since the attractive part of the DD-interaction energy scales as $\sqrt{%
\alpha }$, whereas the kinetic energy in the $x$-direction scales as $\alpha
$, a local minimum can be provided by the interplay between them.

\begin{figure}[tbp]
\centering
\includegraphics[width=0.5\textwidth]{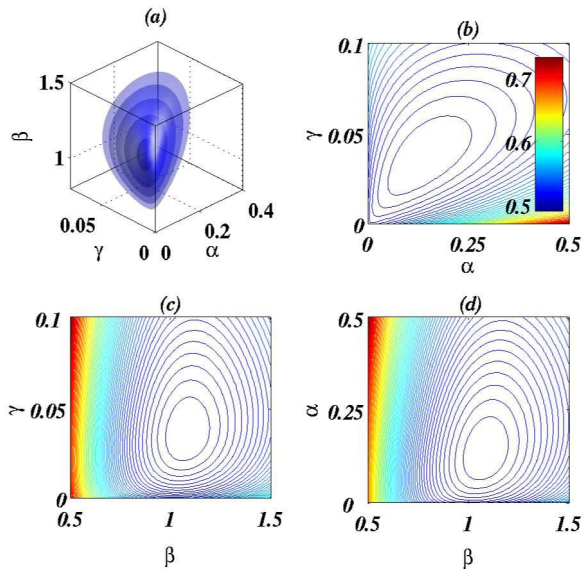}
\caption{(color online) (a) Equal-energy surfaces around the local minimum
of Gross-Pitaevskii energy functional (\protect\ref{enrgpar}), in the
variational parameter space of inverse squared widths $\protect\alpha ,%
\protect\beta ,\protect\gamma $, for $g=10$ and $g/g_{d}=0.911$. Energy
values range from $E=0.4907$ for the innermost surface to $0.4987$ for the
outermost one, in steps of $\Delta E=2\times 10^{-3}$. Panels (b),(c),(d)
depict cross sections through the minimum-energy point, $(\protect\alpha ,%
\protect\beta ,\protect\gamma )_{\mathrm{sol}}=(0.136,1.08,0.037)$, in the $%
\left( \protect\alpha ,\protect\gamma \right) $, $\left( \protect\beta ,%
\protect\gamma \right) $ and $\left( \protect\alpha ,\protect\beta \right) $
planes, respectively. The color bar in (b) applies also to (c) and (d).}
\label{fig2}
\end{figure}

The DD interaction provides for effective attraction at large $L_{x},L_{z}$
and prevents expansion if it asymptotically overcomes the contact
interaction. For $g_{d}>0$, we thus obtain from (\ref{enrgpar}) a threshold
condition for the existence of the soliton, $g/g_{d}<4\pi /3\approx 4.19$.
In Fig \ref{fig2}, we plot equal-energy surfaces and their cross sections in
the $\left( \alpha ,\beta ,\gamma \right) $ space, for $g/g_{d}=0.911$. The
expected local energy minimum exists, corresponding to an elongated soliton
with dimensions $(L_{x},L_{y},L_{z})_{\mathrm{sol}}=(2.71,0.96,5.20)\times
l_{\perp }$. 

\begin{figure}[tbp]
\centering
\includegraphics[width=0.5\textwidth]{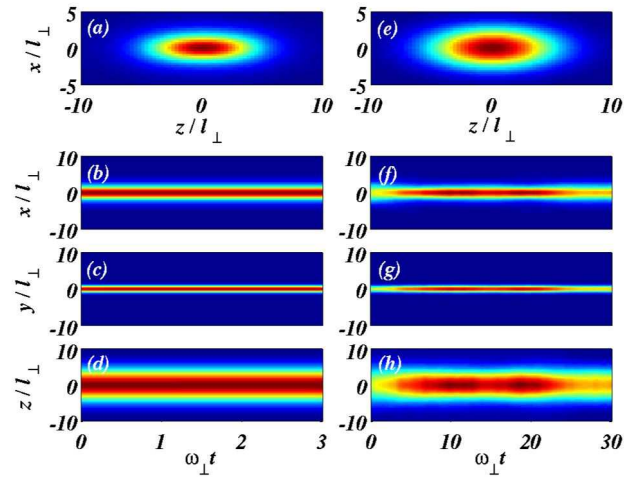}
\caption{(color online) Numerically simulated three-dimensional evolution:
(a) Density of numerically-exact soliton solution at $y=0$, obtained by the
imaginary-time propagation of the Gaussian profile $\protect\psi _{\mathrm{%
aniso}}$ with minimum-energy parameters $\protect\alpha =0.136$, $\protect%
\beta =1.08$, $\protect\gamma =0.037$ borrowed from Fig. \protect\ref{fig2}.
The 2D density profile of this Gaussian function at $y=0$ is shown in (e).
Results of the propagation in real time are presented in (b)-(d) and (f)-(h)
for the initial functions (a) and (e), respectively. Density profiles along
axes $x$,$y$, and $z$, with perpendicular planes going through the soliton's
center, $\left( x_{0},y_{0},z_{0}\right) =(0,0,0)$, are shown as functions
of time. Coefficients $g$ and $g_{d}$ are the same as in Fig. \protect\ref%
{fig2}.}
\label{fig3}
\end{figure}

The existence and stability of the solitons predicted by the above analysis
have been verified by simulations of the underlying 3D GP equation
corresponding to energy functional (\ref{Epar}). The soliton shown in Fig. %
\ref{fig3}(a) was obtained by the 3D propagation in imaginary time of the
Gaussian minimum-energy ansatz depicted in Fig. \ref{fig3}(e). Forward
propagation in real time, starting from this soliton solution (Figs. \ref%
{fig3}(b)-(d)), demonstrates perfect self-focusing. The stability is also
demonstrated by the real-time 3D propagation starting with the Gaussian
approximation, as shown in Figs. \ref{fig3}(f)-(h). The observed long-lived,
slowly damped oscillations indicate the excitation of an internal mode \cite%
{Pelinovsky98}, which is typical to stable solitons in nonintegrable
systems.
%The existence of such modes will be studied elsewhere, within a rigorous stability analysis, based on numerical diagonalization of the Bogoliubov de-Gennes equations around the soliton.

An interesting issue, which will be considered separately, is interactions
between the solitons. Collisions between isotropic solitons supported by the
inverted DD interactions were simulated in Ref. \cite{Pedri05},
demonstrating their fusion into a single soliton. In the present setting,
anisotropy is expected to produce various outcomes of collisions, depending
on the angle between the soliton's polarization and collision direction.

\begin{figure}[tbp]
\centering
\includegraphics[width=0.5\textwidth]{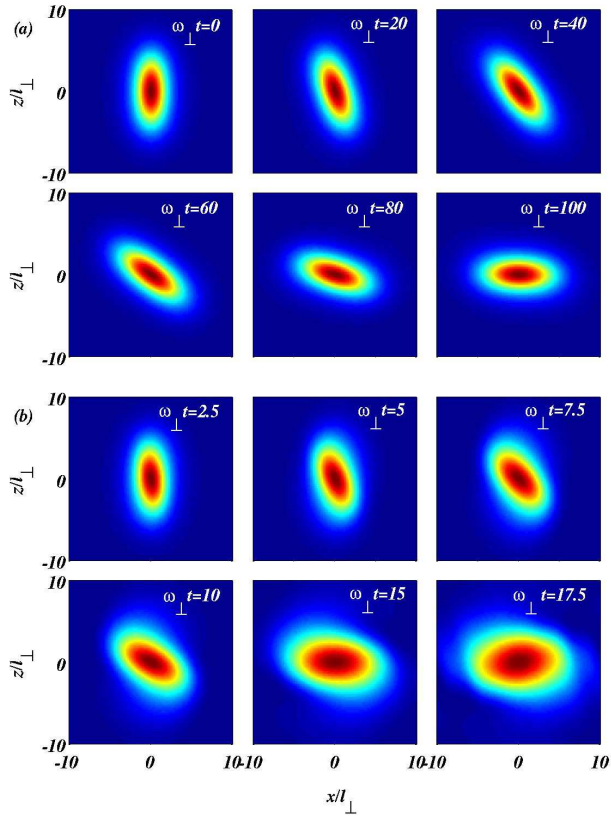}
\caption{(color online) The rotation of the soliton, driven by the
polarization field revolving at angular velocity $\Omega $, starting with
the anisotropic soliton shown in Fig. \protect\ref{fig3}(a), for $\Omega =%
\protect\pi \protect\omega _{\perp }/200$ (a) and $\Omega =\protect\pi
\protect\omega _{\perp }/20$ (b). The interaction strengths are the same as
in Figs. \protect\ref{fig2} and \protect\ref{fig3}.}
\label{fig4}
\end{figure}

Another dynamical behavior, unique to anisotropic solitons, may be attained
by rotation of the external magnetic field used to align the dipole moments.
Whereas the slow rotation of the field is adiabatically followed by the
soliton, as shown in Fig. \ref{fig4}(a), it lags behind faster revolving
fields, see Fig. \ref{fig4}(b). For instance, at $\omega _{\perp }t=10$ the
field is directed along $x$, whereas the long axis of the density profile 
is not parallel to it. This drag is accompanied by significant deformation 
and excitation, as evident from the expansion of the density distribution in 
Fig. \ref{fig4}(b). The amount of excitation around the rotating soliton 
(deviation from adiabaticity) smoothly increases with the rotation velocity, 
roughly like in the case of Landau - Zener tunneling. For slow rotation, it 
is is clearly seen to be exponentially small in the rotation period.

To estimate the experimental feasibility of the proposed quasi-2D solitons
in dipolar BECs, we notice that
%As seen above, the realization of these anisotropic structures does not
%require sophisticated manipulations aimed at the inversion of the sign of
%the DD interaction, unlike the isotropic solitons proposed in Ref. \cite{Pedri05}.
the above-mentioned threshold for the existence of the anisotropic solitons
is $g_{d}/g>0.24$. For $^{52}$Cr with magnetic dipole moment $d=6\mu _{B}$
and $s$-wave scattering length $a\approx 100a_{0}$ ($\mu _{B}$ and $a_{0}$
are the Bohr magneton and radius, respectively), this ratio is $g_{d}/g=\mu
_{0}d^{2}m/(16\pi ^{2}\hbar ^{2}a)=0.036$ \cite{Cr} ($\mu _{0}$ is the
vacuum permeability). Therefore, attenuation of the direct interaction
between atoms may be necessary. A  recent experiment has demonstrated that 
this interaction may be practically switched off by means
of the FR in the condensate of Cr atoms, making the DD interactions
absolutely dominant \cite{recent}. Alternatively,
using BECs of dipolar molecules, with the electric dipole moment on the
order of Debye, will certainly provide sufficiently strong DD interactions.

In conclusion, we have demonstrated, using the variational analysis and
direct simulations of the GP equation in 3D, that the interplay of the
ordinary repulsive contact interactions and long-range DD forces may give
rise to \emph{stable} quasi-2D anisotropic solitons in the dipolar BEC, if
atomic moments are polarized in the 2D plane. Unlike the previously
considered isotropic solitons polarized perpendicular to the plane, the
stability of the anisotropic solitons does not require artificial inversion
of the sign of the dipole-dipole interaction. The anisotropic soliton may
adiabatically follow slow rotation of the polarizing field.

This work was supported in part, by the Israel Science Foundation through a
Center-of-Excellence grant No. 8006/03 and by the Minerva foundation through
a grant for a junior research group.

\end{document}